\documentclass{emulateapj}
\usepackage{amsmath}
\newcommand{\dtau}{{\rm d}\tau}

\newcommand{\Dx}{\Delta x}

\newcommand{\sigmaT}{\sigma_{\rm T}}

\newcommand{\kpch}{h^{-1}{\rm kpc}}
\newcommand{\Mpch}{h^{-1}{\rm Mpc}}

\newcommand{\Ob}{\Omega_{\rm b}}
\newcommand{\Om}{\Omega_{\rm m}}
\newcommand{\Ol}{\Omega_\Lambda}

\newcommand{\Tcmb}{T_{\rm CMB}}

\newcommand{\Aksz}{A_{\rm KSZ}}
\newcommand{\Atsz}{A_{\rm TSZ}}
\newcommand{\Asz}{A_{\rm SZ}}
\newcommand{\Cksz}{C_{\rm KSZ}}
\newcommand{\Ctsz}{C_{\rm TSZ}}
\newcommand{\Csz}{C_{\rm SZ}}
\newcommand{\aksz}{\alpha_{\rm KSZ}}
\newcommand{\atsz}{\alpha_{\rm TSZ}}
\newcommand{\asz}{\alpha_{\rm SZ}}
\newcommand{\muK}{\mu{\rm K}}

\newcommand{\Mfof}{M_{\rm FoF}}
\newcommand{\Mvir}{M_{\rm vir}}
\newcommand{\Rvir}{R_{\rm vir}}

\newcommand{\fgas}{f_{\rm gas}}

\newcommand{\tauT}{\tau_{\rm T}}
\newcommand{\zreion}{z_{\rm reion}}

\newcommand{\Msun}{M_\odot}
\newcommand{\Msunh}{h^{-1}M_\odot}

\shorttitle{Sunyaev-Zel'dovich Angular Power Spectrum}
\shortauthors{Trac, Bode, \& Ostriker}

\begin{document}

\title{Templates for the Sunyaev-Zel'dovich Angular Power Spectrum}

\author{Hy Trac\altaffilmark{1}, Paul Bode\altaffilmark{2}, and Jeremiah P. Ostriker\altaffilmark{2}}
\affil{\altaffilmark{1} Harvard-Smithsonian Center for Astrophysics, Cambridge, MA 02138\\
\altaffilmark{2} Department of Astrophysical Sciences, Princeton University, Princeton, NJ 08544}

\begin{abstract}

We present templates for the Sunyaev-Zel'dovich (SZ) angular power spectrum based on four models for the nonlinear gas distribution. The frequency-dependent SZ temperature fluctuations, with thermal (TSZ) and kinetic (KSZ) contributions, are calculated by tracing through a dark matter simulation, processed to include gas in dark matter halos and in the filamentary intergalactic medium. Different halo gas models are compared to study how star formation, energetic feedback, and nonthermal pressure support influence the angular power spectrum. The standard model has been calibrated to reproduce the stellar and gas fractions and X-ray scaling relations measured from low redshift clusters and groups. The other models illustrate the current theoretical and empirical uncertainties relating to properties of the intracluster medium. Relative to the standard model, their angular power spectra differ by approximately $\pm50\%$ (TSZ), $\pm20\%$ (KSZ), and $\pm40\%$ (SZ at 148 GHz) for $l=3000$, $\sigma_8=0.8$, and homogeneous reionization at $z=10$. The angular power spectrum decreases in amplitude as gas mass and binding energy is removed through star formation, and as gas is pushed out to larger radii by energetic feedback. With nonthermal pressure support, less pressure is required to maintain hydrostatic equilibrium, thus reducing the thermal contribution to the SZ power. We also calculate the SZ templates as a function of $\sigma_8$ and quantify this dependence. Assuming $C_l\propto(\sigma_8/0.8)^\alpha$, the effective scaling index ranges from $7\lesssim\atsz\lesssim9$, $4.5\lesssim\aksz\lesssim5.5$, and $6.5\lesssim\asz{\rm (148\ GHz)}\lesssim8$ at $l=3000$ for $0.6<\sigma_8<1$. The template spectra are publicly available and can be used when fitting for the SZ contribution to the cosmic microwave background on arcminute scales.

\end{abstract}

\keywords{cosmology: theory -- cosmic microwave background -- large-scale-structure of universe -- galaxies: clusters: general -- intergalactic medium -- methods: numerical}

\section{Introduction}

The Sunyaev-Zel'dovich (SZ) effect imprinted in maps of the cosmic microwave background (CMB) is a promising probe of the evolution of large-scale-structure. CMB photons propagating through the expanding universe are scattered by energetic electrons in the intracluster medium (ICM) and intergalactic medium (IGM), resulting in secondary distortions which are the dominant temperature anisotropies on arcminute scales \citep{Sunyaev1970CoASP...2...66S, Sunyaev1972CoASP...4..173S}. Measurements of the frequency-dependent distortions can be used to study the gas distribution in galaxy clusters and groups, and thus to understand the growth of structure. The electron scattering also traces the diffuse baryons in the filamentary cosmic web, which have so far been elusive in detection. Furthermore, the SZ effect has the potential to probe the epoch of reionization when the majority of electrons were dissociated from hydrogen and helium atoms. For reviews, see \citet{Birkinshaw1999PhR...310...97B} and \citet{Carlstrom2002ARA&A..40..643C}.

CMB experiments now have the sensitively and resolution to observe the temperature anisotropies with of order 10$\muK$ noise and arcminute beams. The major SZ science includes the autocorrelation of temperature fluctuations, cross-correlation with large-scale-structure from galaxy surveys, and direct detention of galaxy clusters. In this paper we focus on the SZ angular power spectrum, an interesting statistic because its amplitude depends strongly on the normalization of matter perturbations, generally parameterized by $\sigma_8$. Assuming the SZ power scales as $C_l\propto\sigma_8^\alpha$, previous theoretical calculations find $\atsz\gtrsim7$ \citep[e.g.][]{Seljak2001PhRvD..63f3001S, Komatsu2002MNRAS.336.1256K} for the thermal SZ (TSZ) component, and $\aksz\gtrsim4$ \citep[e.g.][]{Vishniac1987ApJ...322..597V} for the kinetic SZ (KSZ) component. In principle, even an uncertain measurement only within a factor of 2 results in a better than 15\% determination of $\sigma_8$.

Several groups have looked for the SZ contribution to the CMB angular power spectrum beyond the damping tail at multipoles $l\gtrsim1000$. Experiments such as the Arcminute Cosmology Bolometer Array Receiver (ACBAR\footnote{http://cosmology.berkeley.edu/group/swlh/acbar/}), Berkeley-Illinois-Maryland Association (BIMA\footnote{http://bima.astro.umd.edu/}), and Cosmic Background Imager (CBI\footnote{http://www.astro.caltech.edu/$\sim$tjp/CBI/}) report a significant detection of excess power coming from the SZ effect and extragalactic point sources. For example, \citet{Sievers2009arXiv0901.4540S} find an excess that is $1.6\sigma$ above the level expected for $\sigma_8=0.8$ from CBI observations at 30 GHz. However, other experiments such as the Atacama Pathfinder Experiment (APEX-SZ\footnote{http://bolo.berkeley.edu/apexsz/}), Caltech Submillimeter Observatory (Bolocam\footnote{http://www.cso.caltech.edu/bolocam/}), Quest at DASI (QUaD\footnote{http://en.wikipedia.org/wiki/QUaD}) and Sunyaev-Zel'dovich Array (SZA\footnote{http://astro.uchicago.edu/sza/}) report no large excess power.

Most recently, the Atacama Cosmology Telescope (ACT\footnote{http://www.physics.princeton.edu/act/}) and South Pole Telescope (SPT\footnote{http://spt.uchicago.edu/spt/}) have made unprecedented signal-to-noise measurements of the CMB angular power spectrum out to $l\sim10^4$. Both groups fit for the SZ contribution using templates constructed for a $\Lambda$CDM cosmology with $\sigma_8=0.8$, which we presented in \citet{Sehgal2010ApJ...709..920S}. They allow the normalization to vary and measure the scaling factor $\Asz\equiv C_l(\sigma_8)/C_l(0.8)$. \citet{Fowler2010arXiv1001.2934T} place an upper limit (95\% confidence level) of $\Asz < 1.63$ based on ACT observations at 148 GHz. This implies $\sigma_8<0.86$ (95\% CL) if $C_l\propto\sigma_8^7$ is assumed. \citet{Lueker2009arXiv0912.4317L} report a best-fit value of $\Asz=0.42\pm0.21$ (at 153 GHz) for SPT observations near 150 and 220 GHz. When combined with the Wilkinson Microwave Anisotropy Probe (WMAP\footnote{http://lambda.gsfc.nasa.gov/product/wmap/}) 5-year constraints, the joint observations yield $\sigma_8=0.773\pm0.025$.

Currently, there are two publicly available sets of templates for the frequency-dependent SZ angular power spectrum. The \citet[][KS02]{Komatsu2002MNRAS.336.1256K} template\footnote{http://lambda.gsfc.nasa.gov/product/map/dr4/pow\_sz\_spec\_get.cfm} is a popular choice and has been used in most analyses to date. The angular power spectrum is calculated using an analytical halo model, where the gas has a polytropic equation of state and is in hydrostatic equilibrium with a \citet*[][NFW]{Navarro1997ApJ...490..493N} gravitational potential. With thermal but not kinetic contributions to the temperature fluctuations, this template is expected to scale approximately as $C_l\propto\sigma_8^7(\Ob h)^2$. In \citet{Sehgal2010ApJ...709..920S}, we constructed SZ maps\footnote{http://lambda.gsfc.nasa.gov/toolbox/tb\_cmbsim\_ov.cfm} for an octant of the sky by tracing through a dark matter simulation processed to include gas in dark matter halos and in the filamentary IGM. The halo gas distribution is modeled with a polytropic equation of state, and the hydrostatic balance is performed directly on the simulated dark matter halos to preserve the concentration, substructure, and triaxiality of each system \citep{Ostriker2005ApJ...634..964O, Bode2007ApJ...663..139B, Bode2009ApJ...700..989B}. In the standard model of \citet{Bode2009ApJ...700..989B}, star formation and energetic feedback are calibrated against observations of low redshift clusters and groups. These templates have recently been used by ACT and SPT, as discussed above.

In this paper, we address two main questions necessary to interpret observations. How do the SZ temperature fluctuations depend on the assumed model for the nonlinear gas distribution? How does $C_l$ scale with $\sigma_8$ for each given model? We present templates for the SZ angular power spectrum based on four models for the nonlinear gas distribution in a $\Lambda$CDM universe with arbitrary $\sigma_8$. Different halo gas models are compared to study how star formation, energetic feedback, and nonthermal pressure support influence the SZ effect. Section \ref{sec:SZ} reviews the formalism for calculating the SZ temperature fluctuations, and Section \ref{sec:sims} describes the construction of the numerical models. The SZ angular power spectra are compared in Section \ref{sec:results}, and constraints on $\sigma_8$ from recent observations are discussed in Section \ref{sec:discuss}. We adopt the cosmological parameters: $\Om=0.264$, $\Ol=0.736$, $\Ob=0.044$, $h=0.71$, $n_s=0.96$, and $\sigma_8=0.80$, which are similar to the recent WMAP 7-year results \citep{Komatsu2010arXiv1001.4538K}.

\section{Sunyaev-Zel'dovich Effect}
\label{sec:SZ}

The SZ effect is commonly considered to have two main components \citep{Sunyaev1970CoASP...2...66S, Sunyaev1972CoASP...4..173S}. The thermal SZ (TSZ) term arises from inverse Compton scattering of the CMB with hot electrons, predominantly associated with shockheated gas in galaxy clusters and groups. The kinetic SZ (KSZ) term is a Doppler term coming from scattering with electrons having fast, peculiar motions. Another common distinction is that the KSZ effect has a nonlinear component associated with the small-scale ICM, and a more linear component coming from the large-scale IGM. The signal arising from the latter was first calculated for the linear regime by \citet{Ostriker1986ApJ...306L..51O} and \citet{Vishniac1987ApJ...322..597V}, and is often referred to as the OV effect. In this section, we first write down the formalism for the nonrelativistic limit and then the more general relativistic case.

Modeling the SZ effect requires knowing the number density $n_e$, temperature $T_e$, and velocity $v_e$ of the electron distribution. In the nonrelativistic limit, the change in the CMB temperature at frequency $\nu$ in the direction $\hat{n}$ on the sky is given by
\begin{equation}
\frac{\Delta T}{\Tcmb}(\hat{n})  = \left(\frac{\Delta T}{\Tcmb}\right)_{\rm tsz} + \left(\frac{\Delta T}{\Tcmb}\right)_{\rm ksz} = f_\nu y - b,
\label{eqn:SZnr}
\end{equation}
where the dimensionless Compton $y$ and Doppler $b$ parameters,
\begin{gather}
\label{eqn:TSZ}
y \equiv \frac{k_B \sigma_T}{m_e c^2}\int n_e T_e {\rm d}l =  \int \theta_{e}\dtau , \\
 \label{eqn:KSZ}
b \equiv \frac{\sigmaT}{c} \int n_e v_{\rm los}{\rm d}l =  \int \beta_{\rm los}\dtau ,
\end{gather}
are proportional to integrals of the electron pressure and momentum along the line-of-sight (los), respectively. The dimensionless temperature, los peculiar velocity, and the optical depth through a path length ${\rm d}l$ are given by
\begin{gather}
\theta_e \equiv \frac{k_{\rm B}T_e}{m_e c^{2}} = 1.96\times10^{-3} \left(\frac{k_{\rm B}T_e}{{\rm keV}}\right), \\
\beta_{\rm los} \equiv \frac{v_{\rm los}}{c} = 3.34\times10^{-4} \left(\frac{v_{\rm los}}{100\ {\rm km/s}}\right), \\
\dtau \equiv \sigmaT n_e {\rm d}l = 2.05\times10^{-3} \left(\frac{n_e}{10^{-3}\ {\rm cm}^3}\right)\left(\frac{{\rm d}l}{{\rm Mpc}}\right),
\end{gather}
respectively. For a typical cluster, the Compton $y$ parameter is expected to be approximately an order of magnitude larger than the Doppler $b$ parameter.

The nonrelativistic TSZ component has a frequency dependence as specified by the function,
\begin{equation}
f_\nu \equiv x_\nu\coth(x_\nu/2) - 4 , \quad x_\nu\equiv h\nu/(k_{\rm B}\Tcmb) ,
\end{equation}
which has a null at $\nu\approx218$ GHz. The distortion appears as a temperature decrement at lower frequencies and as an increment at higher frequencies relative to the null. The nonrelativistic KSZ component is independent of frequency, but the sign of the distortion depends on the sign of the los velocity. We chose the convention where $v_{\rm los}>0$ if the electrons are moving away from the observer.

In the general, relativistic case, the change in the CMB temperature at frequency $\nu$ in the direction $\hat{n}$ is given by
\begin{align}
\frac{\Delta T}{\Tcmb}(\hat{n}) = \int
\biggl[ & \theta_{e} (Y_0 + \theta_e Y_1 + \theta_e^2 Y_2 + \theta_e^3 Y_3 + \theta_e^4 Y_4) \biggr . \nonumber\\
\biggl . & + \beta^2 \left[\frac{1}{3} Y_0 + \theta_e \left(\frac{5}{6} Y_0 + \frac{2}{3} Y_1\right) \right] \biggr . \nonumber\\
\biggl . & - \beta_{\rm los} (1 + \theta_e C_{1} + \theta_e^2 C_2) \biggr] {\rm d}\tau,
\label{eqn:SZ}
\end{align}
where the $Y$'s and $C$'s are known frequency-dependent coefficients \citep{Nozawa1998ApJ...508...17N}. Note that Eq.\ \ref{eqn:SZnr} contains only the first-order terms in Eq.\ \ref{eqn:SZ} and that $f_\nu = Y_0$.

In this paper, we use Equation (\ref{eqn:SZ}) to calculate the total SZ signal. When the TSZ and KSZ effects are discussed individually, Equations (\ref{eqn:TSZ}) and (\ref{eqn:KSZ}) are used to calculate the Compton $y$ and Doppler $b$, respectively. After multiplying the Compton $y$ by $\Tcmb=2.726\ \mu$K, it becomes equivalent to the nonrelativistic $|\Delta T_{\rm TSZ}|$ at 146 and 280 GHz. The Doppler $b$ temperature fluctuation is equivalent to $\Delta T_{\rm KSZ}$ at all frequencies.

\section{Simulations}
\label{sec:sims}

The SZ effect is modeled by postprocessing a simulation of the large-scale structure of the universe, as described in detail in \citet{Sehgal2010ApJ...709..920S}. In this section, we summarize the methodology and present additional details on the modeling.

The matter distribution along a past light cone, spanning the redshift range $0 \leq z < 10$ and covering one octant of the sky, is generated from an N-body simulation of $1024^3$ dark matter particles evolved
within a periodic box of comoving side length $L = 1000\ \Mpch$. The resolution is set by the particle mass, $m_{\rm p} = 6.82\times10^{10}\ \Msunh$, and the gravitational spline softening length, $\epsilon=16.28\ \kpch$. The friends-of-friends (FoF) halo mass function is in good agreement with the fitting formula from \citet{Jenkins2001MNRAS.321..372J} down to $\Mfof\sim7\times 10^{12}\ \Msunh$ (100 particles).

For lower redshifts $z < 3$, the positions and velocities of all particles in the light cone are saved for postprocessing. The particles thus saved allow a higher-resolution reconstruction of dark matter halos and filaments in the large-scale structure. At higher redshifts $3 \leq z < 10$, projected information is saved instead. Within thin redshift shells, particles are subdivided by angular coordinates and then projected along the line of sight to construct surface density fields for mass and momentum. There are 579 such shells, with thickness $\Delta z\approx 0.09$ at $z\approx 10$ and $\Delta z\approx 0.03$ at $z\approx 3$.

The corresponding gas and electron distributions are modeled with contributions from three components:
\begin{enumerate}

\item Gas in massive dark matter halos with $\Mfof > 2\times 10^{13}\ \Msunh$ at $z<3$ is modeled with a polytropic equation of state and in hydrostatic equilibrium \citep{Ostriker2005ApJ...634..964O, Bode2007ApJ...663..139B, Bode2009ApJ...700..989B}. For each halo, the matter profile is reconstructed at high resolution using the saved N-body particles. This procedure preserves the concentration, substructure, and triaxiality of the system. See Section \ref{sec:gasmodels} for additional discussion.

\item Gas in lower mass halos and the IGM at $z<3$ are modeled using all other saved N-body particles not associated with the massive halos discussed above. The effective gas velocity and temperature are approximated using the peculiar velocity and velocity dispersion of the particles, respectively.

\item At higher redshifts $3 \leq z < 10$, the gas is assumed to trace the matter, and the temperature ($T  \sim 10^4$ K) is predominantly set by photoionization rather than shockheating. The electron distribution is modeled using the saved projected mass and momentum density fields. Hydrogen reionization is assumed to occur instantaneously at $z=10$ and helium is only singly ionized.

\end{enumerate}
Sky maps of the SZ effect are made by tracing through the simulated electron distribution and projecting the accumulated temperature fluctuations onto a HEALPix\footnote{http://healpix.jpl.nasa.gov/} \citep{Gorski2005ApJ...622..759G} grid with pixel resolution of 0.4 arminutes (Nside = 8192). See \citet{Sehgal2010ApJ...709..920S} for additional details and discussion.

\subsection{Halo gas models}
\label{sec:gasmodels}

The hot gas distribution associated with galaxy clusters and groups is modeled as having a polytropic equation of state and being in hydrostatic equilibrium \citep{Ostriker2005ApJ...634..964O, Bode2007ApJ...663..139B, Bode2009ApJ...700..989B}. In this paper, we compare four gas models to study how gas physics influence the SZ effect, in particular the angular power spectrum of the secondary temperature anisotropies:
\begin{enumerate}

\item Adiabatic model, which has neither star formation nor energetic feedback to represent nonradiative gas physics.

\item Standard model, with star formation and feedback calibrated against observations of nearby clusters and groups.

\item Low-$\fgas$ model, with twice the star formation and twice the feedback energy input as the standard model.

\item Nonthermal20 model, which includes 20\% nonthermal pressure support plus more star formation, but less feedback energy input, than the standard model.

\end{enumerate}

\begin{figure}[t]
\center
\includegraphics[width=3.5in]{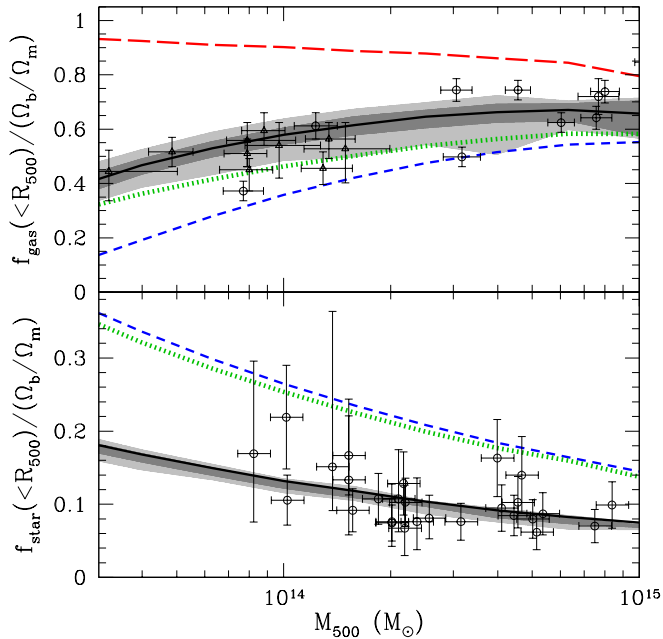}
\caption{{\it Top}: Gas fractions for halos at $z<0.2$ from the standard (black solid), adiabatic (red long-dash), low-$\fgas$ (blue short-dash), and nonthermal20 (green dotted) models. The $1\sigma$ (dark gray) and $2\sigma$ (light gray) scatters from the standard model are also shown. Observational data points are from \citet[circles]{Vikhlinin2006ApJ...640..691V} and \citet[triangles]{Sun2009ApJ...693.1142S}. {\it Bottom}: Stellar fractions for the same halos as above. The standard model adopts the best-fit relation from \citet[circles]{Lin2003ApJ...591..749L}, the low-$\fgas$ model has twice the stellar fraction as the standard model, while the nonthermal20 model is taken from \citet{Giodini2009ApJ...703..982G}.}
\label{fig:fgasfstar}
\end{figure}

\begin{figure*}[t]
\center
\includegraphics[width=3.5in]{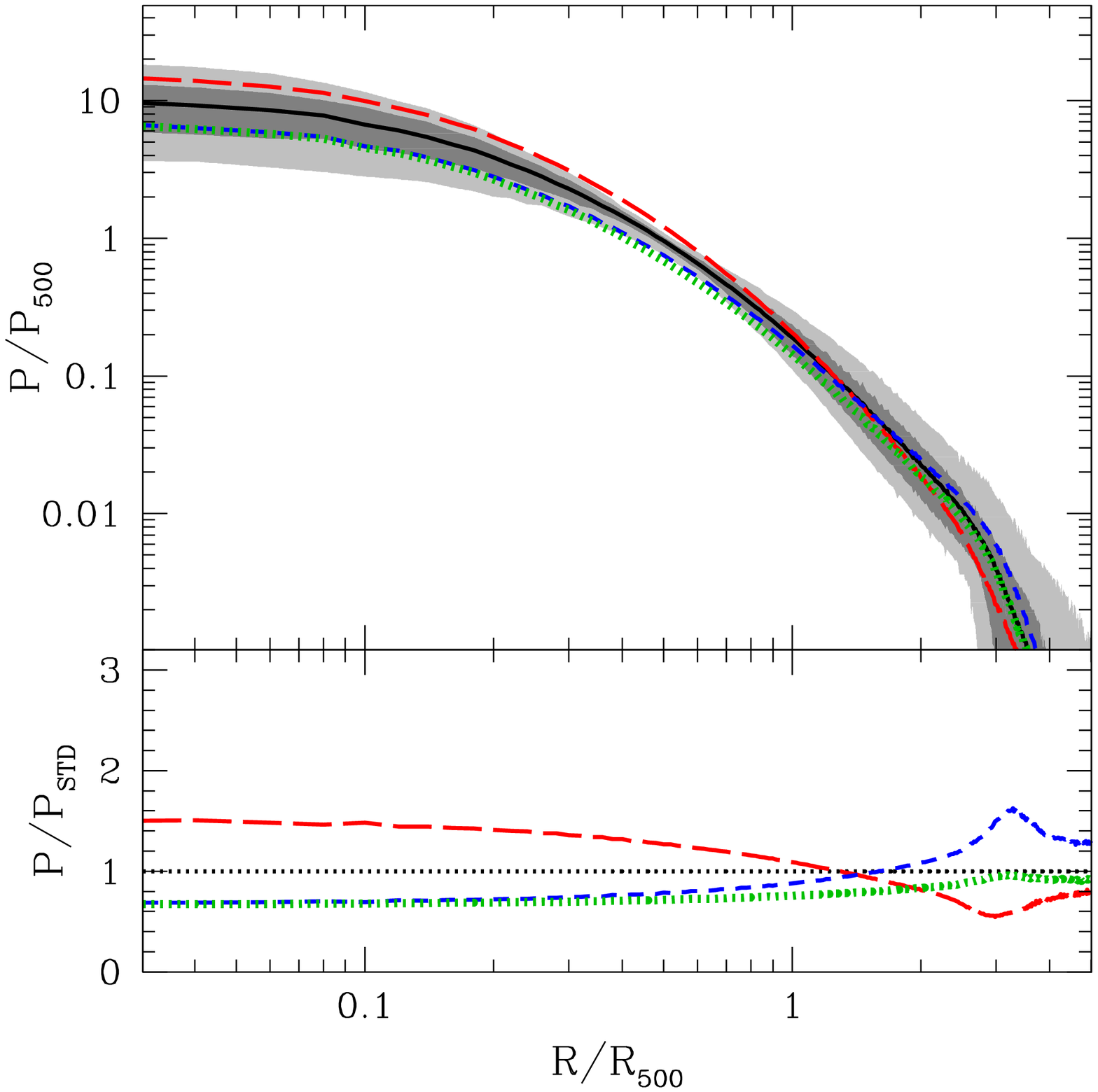}\includegraphics[width=3.5in]{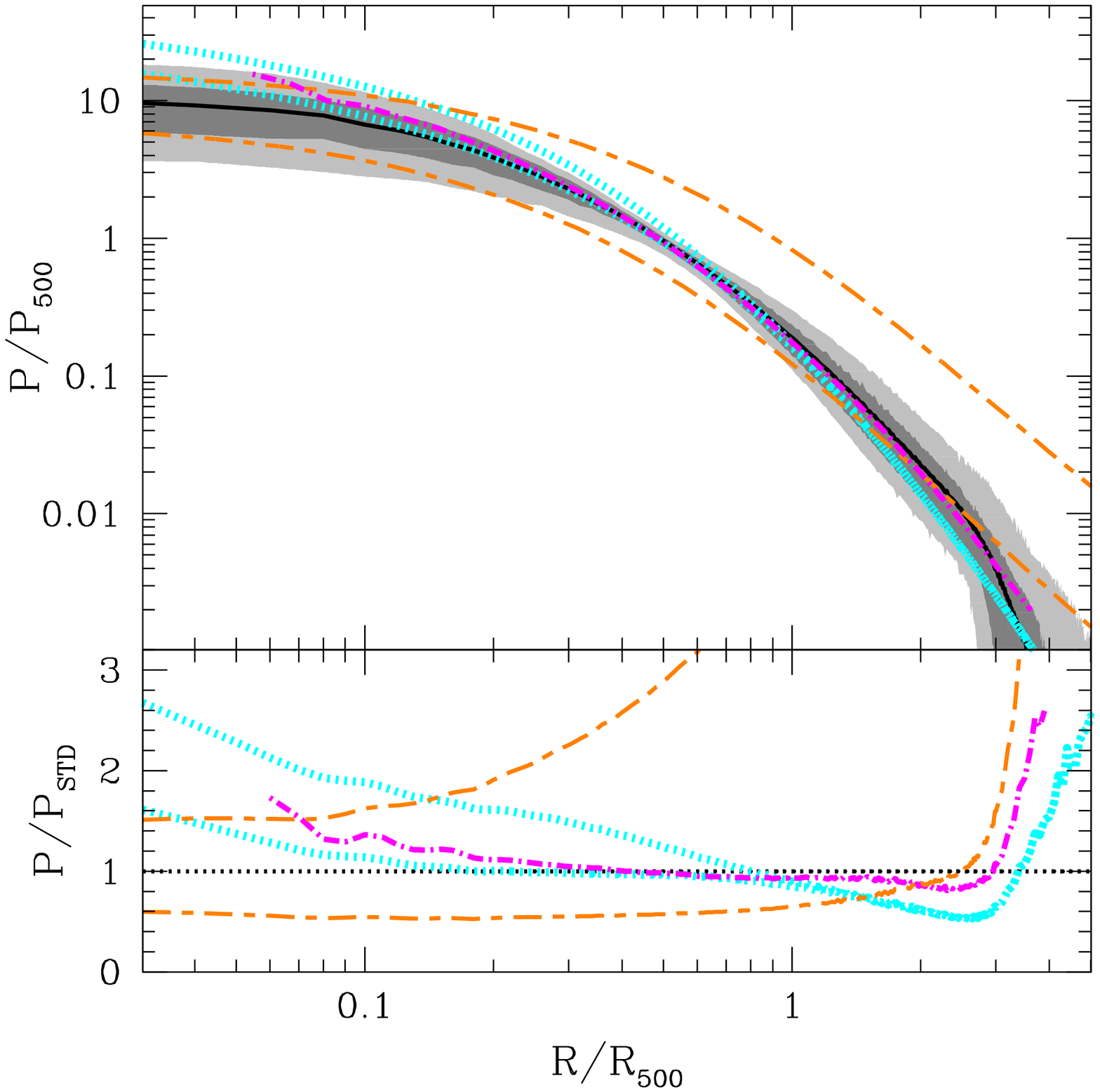}
\caption{{\it Left}: Average pressure profiles for halos with $M_{500}>10^{14}\ \Msun$ and $z<0.2$ from the standard model (black solid), adiabatic (red long-dash), low-$\fgas$ (blue short-dash), and nonthermal20 (green dotted) models. The $1\sigma$ (dark gray) and $2\sigma$ (light gray) scatters from the standard model are also shown. The bottom panel shows pressure ratios with respect to the standard model. Note a kink at $R/R_{500}\sim3$ is present because the IGM contribution is not included here for the simulated profiles. {\it Right}: The \citet[cyan dotted]{Arnaud2009arXiv0910.1234A} and \citet[orange short-long-dash]{Komatsu2001MNRAS.327.1353K, Komatsu2002MNRAS.336.1256K} profiles are for $M_{500} = 10^{14}\ \Msun$ (lower) and $10^{15}\ \Msun$ (upper) at $z=0.2$. The \citet[magenta dot-dash] {Battaglia2010arXiv1003.4256B} profile is a weighted average from simulated halos in the mass range $10^{14}< M_{500}/\Msun < 10^{15}$ at $z=0$.}
\label{fig:p_of_r}
\end{figure*}

Figure \ref{fig:fgasfstar} shows the stellar mass fraction $f_*$ and gas mass fraction $\fgas$ within $R_{500}$ at $z<0.2$ for these models. The mass fractions are defined relative to $M_{500}$ and normalized by the cosmic baryon fraction $\Ob/\Om$. The adiabatic model provides an appropriate upper limit on the gas fraction. In the absence of star formation and feedback, the gas fractions are close to the cosmic average and varies only weakly with mass and redshift. The slight increase towards lower masses is due to the higher halo concentrations. While this basic model is known to be incorrect, it is still useful in comparison for understanding the effects of radiative gas physics.

\citet{Bode2009ApJ...700..989B} previously constructed the standard model to match the stellar and gas fractions observed in nearby  clusters and groups. The stellar fractions at $z=0$ are given by $f_*(<R_{500})=0.0164(M_{500}/3\times10^{14}\ \Msun)^{-0.26}$ \citep{Lin2003ApJ...591..749L}, and the redshift evolution follows a delayed exponential model \citep{Nagamine2006ApJ...653..881N}. In order to match the gas fractions from X-ray observations \citep{Vikhlinin2006ApJ...640..691V, Sun2009ApJ...693.1142S}, a feedback efficiency $\epsilon=4\times 10^{-6}$ and energy input of $\epsilon M_{\rm F}c^2$ is required, where $M_{\rm F}$ is the mass of formed stars. With this calibration, the standard model reproduces several observed X-ray scaling relations, such as $T_X-M_{500}$, $L_X-M_{500}$, and $Y_X-M_{500}$. We choose the standard model as the fiducial case when comparing with other models and with observations.

The low-$\fgas$ model has even higher stellar fractions and lower gas fractions than the standard model. This model has twice as much star formation and twice as much feedback energy input. The feedback efficiency constant is the same for both models, and for the adopted value we find that star formation and feedback have comparable effects on reducing the gas fraction within $R_{500}$. The gas fractions skirt the lower bounds of the uncertainty in the measurements from nearby clusters and groups. While the $T_X-M_{500}$ relation for $M_{500}\gtrsim10^{14}\ \Msun$ is in good agreement with observations, the $L_X$ and $Y_X$ values are lower by approximately 20\%. The low-$\fgas$ model will help in understanding what is a reasonable lower bound on the SZ angular power spectrum due to uncertainty in the star formation and feedback histories.

The nonthermal20 model is different from the other three models in that 20\% of the hydrostatic pressure is assumed to be nonthermal. Turbulent and relativistic contributions account for 15\% and 5\%, respectively, of the pressure at all radii. Note these two components contribute differently to the total
energy budget, because of the different ratios of specific heats. This model also includes star formation and feedback, but different in detail than previously discussed. The stellar fraction follows Equation 10 of \citet{Giodini2009ApJ...703..982G}: $f_*(<R_{500})=0.05(M_{500}/5\times10^{13}\ \Msun)^{-0.26}$. This is a fit to X-ray-selected groups and poor clusters of lower mass than the \citet{Lin2003ApJ...591..749L} sample;  as it turns out, this is almost twice as large as the $f_*$ used in the standard model. A lower feedback efficiency, $\epsilon=1\times 10^{-6}$, is needed to reproduce the gas fractions and scaling relations from X-ray observations. The model does this well, keeping in mind that the total masses derived observationally must also be corrected for the 20\% nonthermal contribution in the hydrostatic equilibrium equation.

There is observational support for including both turbulent/bulk motions and cosmic rays. Comparing X-ray and weak lensing masses, \citet{Zhang2010ApJ...711.1033Z} find nonthermal pressure support of $\sim 9\%$ inside $R_{500}$, while \citet{Mahdavi2008MNRAS.384.1567M} found roughly twice this amount. Comparing the gravitational potential profiles of the central galaxies of the Fornax and Virgo clusters (derived from X-ray and from optical data), \citet{Churazov2008MNRAS.388.1062C} find the total nonthermal pressure to be $\lesssim 10-20\%$ of the gas thermal pressure. Examining the width of X-ray emission lines, \citet{Sanders2010MNRAS.402L..11S} place an upper limit of 13\% on the turbulent energy density as a fraction of the thermal energy density in the core of A1835. Simulations of clusters give values for energy in cosmic rays relative to the thermal energy of approximately $5-10\%$ \citep[e.g.][]{Jubelgas2008A&A...481...33J}, in good agreement with constraints from gamma ray and radio observations \citep[e.g.][]{Pfrommer2004A&A...413...17P}. Simulations likewise give contributions from bulk flows and turbulence of approximately $5-20\%$ \citep[e.g.][]{Lau2009ApJ...705.1129L, Meneghetti2009arXiv0912.1343M, Burns2010arXiv1004.3553B}. Thus, to examine the effects of these components we include a generous 20\% nonthermal pressure support in the nonthermal20 model to examine the effects of these components, with turbulence being the dominant contributor.

Figure \ref{fig:p_of_r} shows the radial pressure profiles for halos with $M_{500}>10^{14}\ \Msun$ and $z<0.2$ drawn from our models, and compares them with other theoretical and observational results. We rescale the halo profiles by the characteristic pressure $P_{500}$ \citep[see][]{Nagai2007ApJ...655...98N} and calculate an average by weighting each halo by the integrated pressure within $R_{500}$ \citep[see][]{Battaglia2010arXiv1003.4256B}. At smaller radii ($R\lesssim 1.5R_{500}$), the differences in our models are due to the depletion of gas mass and binding energy through star formation, the redistribution of gas to larger radii through energetic feedback, and the reduction in the amount of thermal pressure needed to maintain hydrostatic equilibrium in the presence of a nonthermal component. At larger radii ($R\gtrsim 1.5R_{500}$), halos experiencing more energetic feedback have shallower profiles as gas gets pushed outwards. When considering the total pressure for $R\lesssim \Rvir \approx 2R_{500}$, the differences between the standard, adiabatic, and low-$\fgas$ models are almost entirely due to the different stellar fractions. In the case of the nonthermal20 model, there is an additional reduction due to the 20\% nonthermal hydrostatic support. The pressure profiles rapidly decrease at $R/R_{500}\gtrsim3$, as the IGM contribution is not included in the plot. The gas in each halo has a different, anisotropic outer limit, which also changes with the model \citep[see][]{Sehgal2010ApJ...709..920S}.

\citet{Arnaud2009arXiv0910.1234A} used the REXCESS cluster sample \citep{Bohringer2007A&A...469..363B} at $z<0.2$ and found that the scaled pressure profiles are well fit by a generalized NFW pressure model \citep{Nagai2007ApJ...655...98N}. The pressure profiles from the standard model are in good agreement with this recent analysis. At small radii ($R/R_{500} < 0.1$), we underpredict the pressure because of the finite resolution of the gas reconstruction, which is computed using a Cartesian grid with cell length $\Dx=32.55\ \kpch\ ({\rm comoving}) \approx 0.03 R_{500}$. However, this small region contributes only minorly to the total pressure. At intermediate radii ($0.1 < R/R_{500} < 1$), which is the most relevant range for this comparison, our profiles are similar to the REXCESS sample \citep[also see Figure 8 in][]{Arnaud2009arXiv0910.1234A}. Note that on larger radii ($1 < R/R_{500} < 4$), their best-fit model is based on results from hydrodynamic simulations \citep{Nagai2007ApJ...655...98N} rather than X-ray data.

The \citet[KS02]{Komatsu2001MNRAS.327.1353K, Komatsu2002MNRAS.336.1256K} pressure profiles shown in Figure \ref{fig:p_of_r} are for $M_{500} = 10^{14}\ \Msun$ and $10^{15}\ \Msun$ at $z=0.2$ (courtesy of E.~Komatsu). While their gas model is also based on a polytropic gas in hydrostatic equilibrium, there are some important differences to note. Their model is based on nonradiative gas physics (having no stars nor feedback), but the pressure profiles differ from our adiabatic model for the following reasons. First, the KS02 model is normalized such that the baryon fraction at the virial radius is equal to the cosmic average, which has the effect of setting the baryon fraction within this radius to be less than the cosmic average. This is more similar to the standard model, rather than the adiabatic. Second, the polytropic index depends on the halo concentration and is approximately 1.1 for the mass ($M_{500} > 10^{14}\Msun$) and redshift ($z<0.2$) ranges being considered in Figure \ref{fig:p_of_r}, which is lower than our constant value of 1.2. As a result, the KS02 pressure profiles for clusters are shallower and do not decrease as fast with radii.

\citet{Battaglia2010arXiv1003.4256B} recently modeled the SZ effect using hydrodynamic simulations with radiative cooling, star formation, and feedback from supernovae and active galactic nuclei (AGN). The simulations directly capture bulk and turbulent gas motions, and also include a prescription for cosmic rays \citep{Pfrommer2007MNRAS.378..385P}. In Figure \ref{fig:p_of_r}, their scaled pressure profile (courtesy of N.~Battaglia) is a weighted average from halos in the mass range $10^{14}< M_{500}/\Msun < 10^{15}$ at $z=0$. We find very good agreement with our standard model, within $\sim20\%$ (which is smaller than the 1$\sigma$ scatter) over the radial range $0.1 < R/R_{500} < 3$. The nonthermal20 model, which also accounts for cosmic rays and turbulent gas motions, compares even more favorably with theirs at $R/R_{500} > 1$, but less well at smaller radii.

\section{Results}
\label{sec:results}

\begin{figure*}
\center
\includegraphics[width=3.5in]{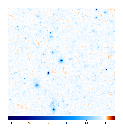}\hspace{1mm}\includegraphics[width=3.5in]{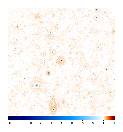}\\
\includegraphics[width=3.5in]{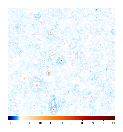}\hspace{1mm}\includegraphics[width=3.5in]{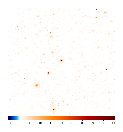}
\caption{{\it Top left}: The SZ temperature anisotropies in $\mu$K at 148 GHz from a $4^\circ \times4^\circ$ sample field in the standard model map. At this frequency, the TSZ signal is a temperature decrement, but the KSZ signal is a decrement (increment) if the los peculiar velocity is positive (negative). The cluster at the center of the field has mass $\Mvir\approx10^{15}\Msun$ and redshift $z\approx0.3$. {\it Top right}: The temperature difference, $\Delta T_{\rm adiabatic}-\Delta T_{\rm standard}$, between the adiabatic and standard models. In the adiabatic model, the gas is more centrally concentrated, leading to relatively stronger central decrements, but slightly weaker decrements at larger radii. {\it Bottom left}: The temperature difference, $\Delta T_{{\rm low}-\fgas}-\Delta T_{\rm standard}$, between the low-$\fgas$ and standard models. With more star formation and feedback, the low-$\fgas$ model has weaker central decrements, but slightly stronger decrements at larger radii. {\it Bottom right}: The temperature difference, $\Delta T_{{\rm nonthermal20}}-\Delta T_{\rm standard}$, between the nonthermal20 and standard models. The nonthermal20 model has less pressure and weaker decrements at all radii.}
\label{fig:maps}
\end{figure*}

\begin{figure}[t!]
\center
\includegraphics[width=\hsize]{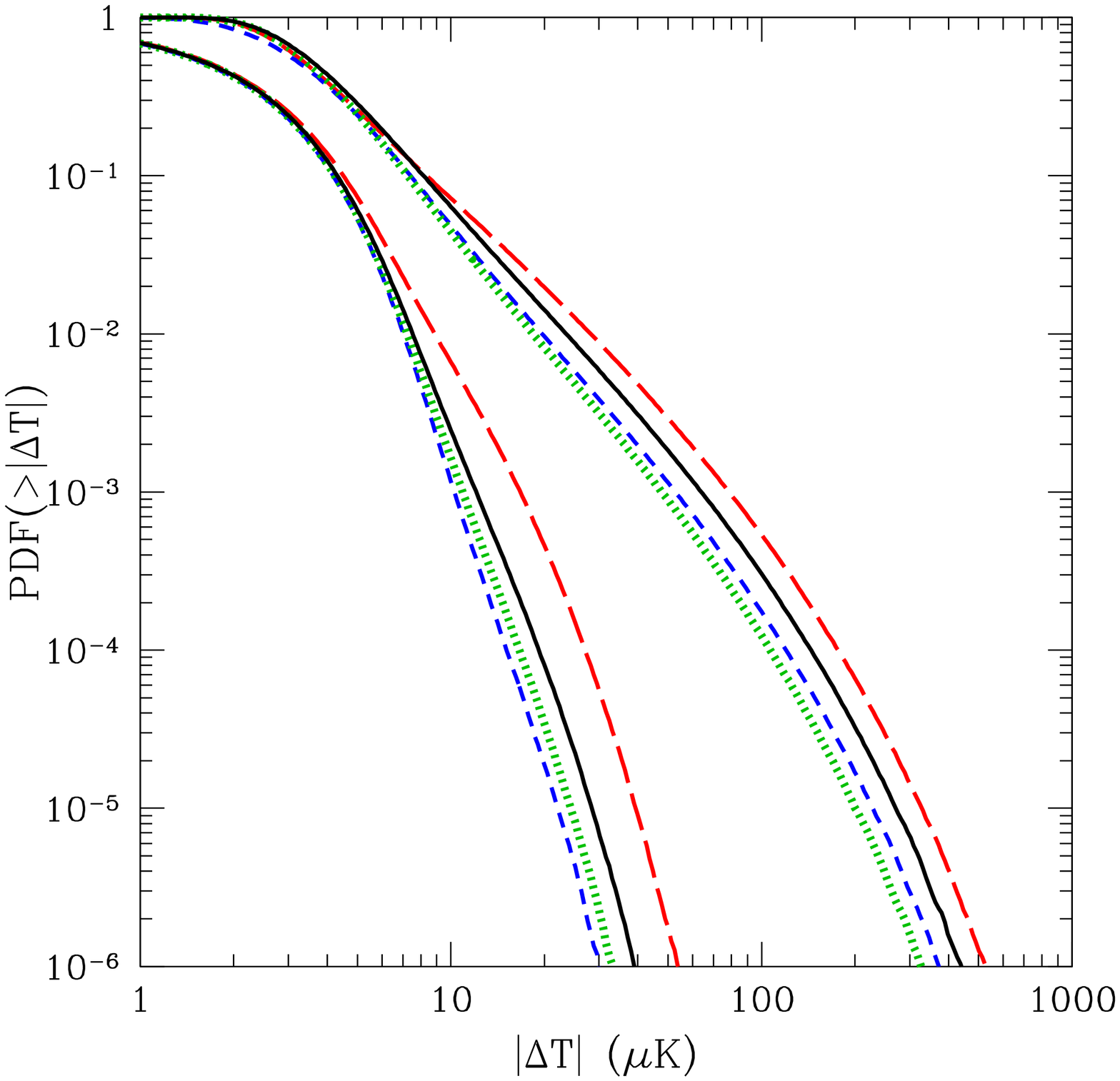}
\caption{Cumulative probability distribution functions of Compton $y$ (right) and Doppler $b$ (left) temperature fluctuations for the standard (black solid), adiabatic (red long-dash), low-$\fgas$ (blue short-dash), and nonthermal20 (green dotted) models. The $y$ and $b$ values have been multiplied by $\Tcmb=2.726\ \mu$K to get the temperature fluctuations in $\mu$K. The angular smoothing scale is set by the map pixel resolution of 0.4 arcminutes.}
\label{fig:pdf}
\end{figure}

We make sky maps of the SZ effect in Healpix format with pixel resolution of 0.4 arcminutes (Nside = 8192). Figure \ref{fig:maps} shows a 4 degree $\times$ 4 degree sample field, centered on a cluster with $\Mvir\approx10^{15}\Msun$ at $z\approx0.3$. The top-left panel shows the SZ temperature anisotropies in $\mu$K at 148 GHz from the standard model, while the other panels show the temperature differences between the standard model and the other three models for the same sample field. At this frequency, the TSZ component is always a temperature decrement, but the KSZ component can be positive or negative.

For comparison, the reference cluster in the middle of the field has a central decrement of approximately 355, 280, 225, and 200 $\mu$K in the adiabatic, standard, low-$\fgas$, and nonthermal20 models, respectively. Comparing the first three models, we find that the central decrement gets weaker as more gas is removed through star formation and as more gas is pushed out to larger radii by energetic feedback. The latter also results in relatively stronger decrements at larger radii. Compared to the standard model, the larger stellar fraction and the 20\% nonthermal pressure support in the nonthermal20 result in less pressure and weaker decrements at all radii. In Sections \ref{sec:tsz} - \ref{sec:sz}, we show that the effects of star formation, energetic feedback, and nonthermal hydrostatic support lead to scale-dependent changes in the power spectrum.

Figure \ref{fig:pdf} compares the cumulative probability distribution functions (PDF) of Compton $y$ and Doppler $b$ temperature anisotropies from maps with pixel resolution of 0.4 arcminutes. After multiplying the values by $\Tcmb=2.726\ \mu$K, the Compton $y$ becomes equivalent to $|\Delta T_{\rm TSZ}|$ at 146 and 280 GHz, and the Doppler $b$ equivalent to $\Delta T_{\rm KSZ}$ at all frequencies. On average, the $|b|$ temperature fluctuations are approximately an order of magnitude smaller than $y$. In principle, the PDF of the SZ signal can be used in place of the halo mass function to constrain cosmological models. However, it will be difficult to break the degeneracy between cosmological and astrophysical parameters, for example between $\sigma_8$ and $f_{\rm gas}(M,z)$, using this statistic alone. Furthermore, it will be difficult to isolate the SZ one-point signal in the presence of the CMB, extragalactic point sources, galactic dust, and noise.

\subsection{TSZ angular power spectra}
\label{sec:tsz}

\begin{figure}[t]
\center
\includegraphics[width=\hsize]{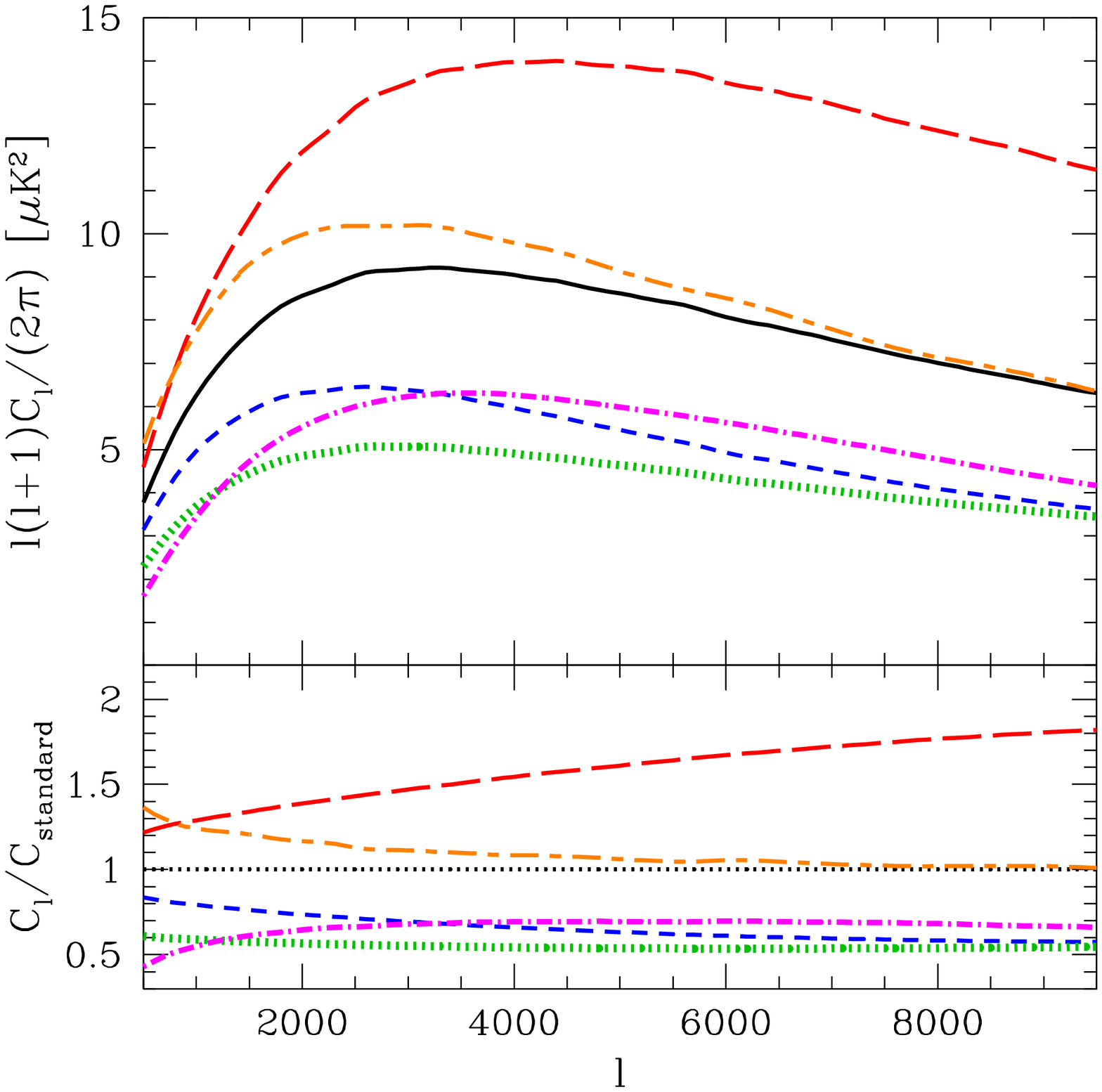}
\caption{{\it Top}: Compton $y$ or TSZ (at 146 and 280 GHz) angular power spectra for the standard (black solid), adiabatic (red long-dash), low-$\fgas$ (blue short-dash), and nonthermal20 (green dotted) models. For comparison, the \citet[orange short-long-dashes]{Komatsu2002MNRAS.336.1256K} and \citet[magenta dash-dot]{Battaglia2010arXiv1003.4256B} templates have been converted to the same frequency. {\it Bottom}: Ratios of power spectra with respect to the standard model. Although the KS02 template is based on nonradiative gas physics, it resembles the standard model much more than the adiabatic model.}
\label{fig:cl_tsz}
\end{figure}

\begin{figure*}[t]
\center
\includegraphics[width=3.5in]{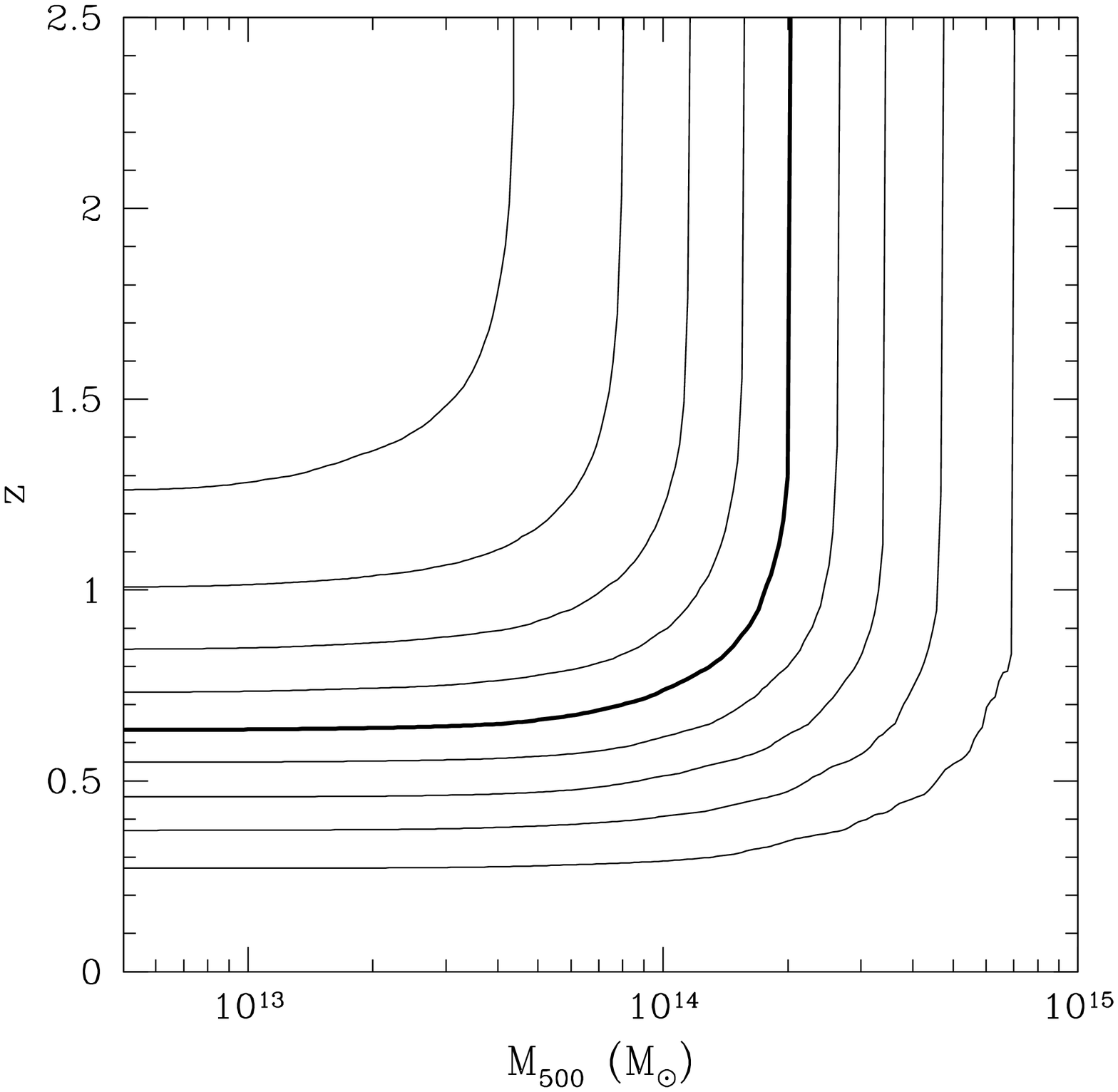}\includegraphics[width=3.5in]{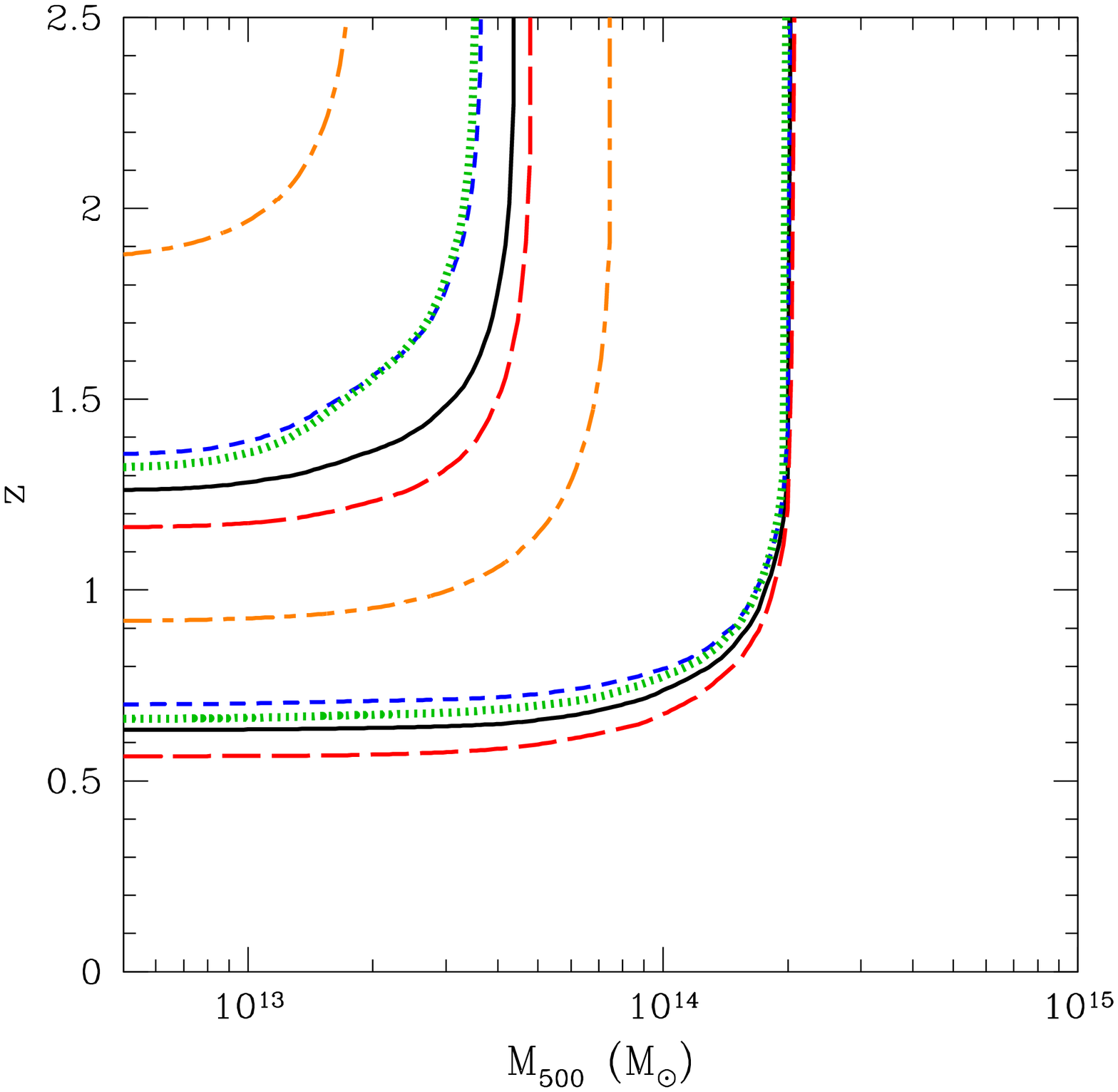}
\caption{{\it Left:} The fraction of the thermal power (from both halos and the IGM) at $l=3000$ as a function of minimum mass and maximum redshift in the standard model (black solid). The nine curves (from bottom to top) are for PDF($>M_{500},<z$) = [0.1, 0.9] in increments of 0.1. {\it Right:} PDF($>M_{500},<z$) = 0.5 (bottom) and 0.9 (top) from the adiabatic (red long-dash), low-$\fgas$ (blue short-dash), nonthermal20 (green dotted), and KS02 (orange short-long-dashes) models are compared to the standard model.}
\label{fig:pdf_cl_tsz}
\end{figure*}

Angular power spectra of the Compton $y$ (and equivalently the TSZ at 146 and 280 GHz) temperature fluctuations are shown in Figure \ref{fig:cl_tsz}. The power spectra are calculated from the simulated maps using the software PolSpice\footnote{http://www.planck.fr/article141.html} \citep[e.g.][]{Szapudi2001ApJ...548L.115S, Chon2004MNRAS.350..914C} and have been corrected for the pixel window. In all four models, the TSZ power spectrum is broadly peaked at $l\sim3000$, where it also becomes dominant over the lensed CMB power spectrum.

Relative to the standard model, the adiabatic model is 47\% larger, the low-$\fgas$ model is 30\% smaller, and the nonthermal20 model is 45\% smaller at $l=3000$. The effects of star formation and feedback can be simply understood. As more gas mass and binding energy is removed through star formation and as more gas is pushed out to larger radii by energetic feedback (see Figure \ref{fig:maps}), the overall amplitude of the TSZ angular power spectrum decreases. Since feedback more strongly affects lower mass halos because of their shallower potential wells, there is a stronger reduction in power on smaller angular scales. Correspondingly, as more massive or lower redshift halos make a relatively larger contribution to the total power, the peak in the spectrum shifts to larger angular scales.

Interestingly, we find that the nonthermal20 model has even lower power than the low-$\fgas$ model at all angular scales, despite having higher gas fractions. Since their stellar fractions are very similar, the removal of gas mass and binding energy through star formation is very close to equal in these two models, and thus not responsible for the differences. The two reasons for the differences in the power spectra are as follows. First, while energetic feedback in the low-$\fgas$ model redistributes the mass and energy to larger radii, the nonthermal hydrostatic support in the nonthermal20 model simply reduces the pressure at all radii (see Figure \ref{fig:maps}). Second, the feedback energy per baryon is correlated with the stellar fraction, which decreases at higher redshifts, while the assumed 20\% nonthermal support is independent of $z$. For reasonable values of parameters, nonthermal support can have comparable effects as energetic feedback on lowering the amplitude of the thermal power. Also, see \citet{Shaw2010arXiv1006.1945S} for recent work using halo model calculations to study the impact of cluster physics on the TSZ angular power spectrum.

Figure \ref{fig:pdf_cl_tsz} shows the fraction of the thermal power (from both halos and the IGM) at $l=3000$ as a function of minimum mass and maximum redshift, denoted by PDF($>M_{500},<z$). For the standard model, all halos with $M_{500}\gtrsim1.5\times 10^{14}\ (3\times 10^{13})\ \Msun$ and $z\lesssim0.85$ (1.5) contribute approximately 50\% (90\%) of the power. The other three models require similar mass and redshift ranges, but there are systematic differences. Models having higher stellar fractions at low redshifts (e.g.~low-$\fgas$ and nonthermal20 models) have relatively larger contributions coming from higher redshifts as $f_*$ declines. Also, models with gas fractions which decrease faster toward low masses (e.g.~low-$\fgas$) have relatively larger fractions of the power coming from massive halos with $M_{500} \gtrsim 3\times10^{14}\ \Msun$. Figure \ref{fig:pdf_cl_tsz} demonstrates that in order to build templates for the TSZ angular power spectrum accurate to 10\%, we need to calibrate theoretical pressure profiles using a fair sample of clusters and groups out to $z\sim1.5$.

Overall, the four models are similar to results from hydrodynamic simulations \citep[e.g.][]{Refregier2000PhRvD..61l3001R, Seljak2001PhRvD..63f3001S, Springel2001ApJ...549..681S, daSilva2001MNRAS.326..155D, Bond2002ASPC..257...15B, Bond2005ApJ...626...12B, Hallman2007ApJ...671...27H, Battaglia2010arXiv1003.4256B} and semi-analytical/numerical calculations \citep[e.g.][]{Komatsu2002MNRAS.336.1256K, Holder2007MNRAS.382.1697H, Moodley2009ApJ...697.1392M, Peel2009MNRAS.397.2189P, Shaw2010arXiv1006.1945S}. However, there are still factor of $\sim2$ differences amongst the various predictions, even after correcting for different cosmologies using the approximate scaling $C_l\propto\sigma_8^7(\Ob h)^2$ \citep{Seljak2001PhRvD..63f3001S, Komatsu2002MNRAS.336.1256K}. It is generally difficult to correct for different astrophysical assumptions relating to star formation and feedback. Note that previous hydrodynamic simulations are often limited to small box sizes, where the abundance of clusters (and even groups) are underpredicted due to missing large-scale power in the initial conditions. Thus, these results are highly prone to sample variance.

For comparison, the popular KS02 template is also shown in Figure \ref{fig:cl_tsz}. It has been converted to the same frequency, but no correction for cosmological parameters is applied, because $\sigma_8=0.8$ is the same and $(\Ob h)^2$ is larger by only 4\% compared to our cosmology.  At $l=3000$, it has approximately 10\% more power than the standard model. Although the KS02 model has no stars and feedback, it resembles the standard model much more than the adiabatic model. This is due to the choice of density normalization and polytropic indices, as previously explained in Section \ref{sec:gasmodels}. The KS02 template also has a slightly different shape, with relatively more power on larger angular scales for the following reasons. First, the massive halos have lower polytropic indices and more extended pressure profiles. Second, when calculating the angular power spectrum with the Limber equation \citep[see Equations (1)-(2) in][]{Komatsu2002MNRAS.336.1256K}, the pressure profiles are integrated out to $\gtrsim 3\Rvir$ to improve numerical convergence. Figure \ref{fig:pdf_cl_tsz} shows that all halos with $M_{500}\gtrsim5\times 10^{13}\ (1\times 10^{13})\ \Msun$ and $z\lesssim1.1$ (2.0) contribute approximately 50\% (90\%) of the thermal power at $l=3000$ in the KS02 template. In their gas model, the concentration parameter and polytropic index decrease toward higher redshift at fixed mass, which enhances the pressure at $R/R_{500} \gtrsim 0.5$. Consequently, high redshift and low mass halos contribute relatively more to the power than in any of our models. 

\citet{Battaglia2010arXiv1003.4256B} have calculated the TSZ angular power spectra from many realizations of maps (maximum size $= 3.2^\circ\times3.2^\circ$) made with hydrodynamic simulations (maximum box size $= 330\ \Mpch$). In Figure \ref{fig:cl_tsz} we show their fiducial case based on a gas model with AGN feedback (courtesy of N.~Battaglia). No correction for cosmological parameters is applied,  since $\sigma_8=0.8$ is the same and $(\Ob h)^2$ is only 2\% smaller than our value. At $l=3000$, this case has approximately 32\% less power than the standard model, but 23\% more power than the nonthermal20 model. These differences are consistent with those found in the average pressure profile of low redshift clusters (see Section \ref{sec:gasmodels}), which is reassuring. However, it is not conclusive that the two approaches would yield similar power spectra for the same low redshift pressure model since the majority of the power comes from higher redshifts. Different implementations of gas prescriptions may not have the same evolution of gas and stellar properties. Furthermore, the finite box and map sizes may have different amounts of sample variance, in particular at larger angular scales. A more detailed comparison, for example using the PDF($>M_{500},<z$), is required in order to check the robustness of agreement in various approaches.

\subsection{KSZ angular power spectra}
\label{sec:ksz}

\begin{figure}[t]
\center
\includegraphics[width=\hsize]{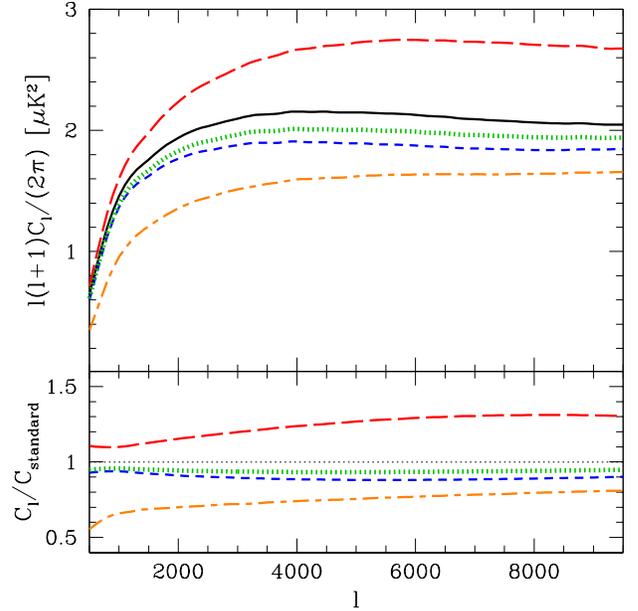}
\caption{{\it Top}: KSZ angular power spectra for the standard (black solid), adiabatic (red long-dash), low-$\fgas$ (blue short-dash), and nonthermal20 (green dotted) models, in which hydrogen is homogeneously reionized at $\zreion=10$. For comparison, the standard model is modified to have hydrogen reionized later at $\zreion=6$ (orange short-long-dash). {\it Bottom}: Ratios of power spectra with respect to the standard model.}
\label{fig:cl_ksz}
\end{figure}

Angular power spectra of the KSZ temperature fluctuations are shown in Figure \ref{fig:cl_ksz}.  The KSZ signal is independent of frequency and its angular power spectrum is much broader, with a peak amplitude that is approximately several times smaller compared to the Compton $y$ angular power spectrum. This is consistent with results from hydrodynamic simulations \citep[e.g.][]{Springel2001ApJ...549..681S, daSilva2001MNRAS.326..155D, Zhang2004MNRAS.347.1224Z, Hallman2009ApJ...698.1795H, Battaglia2010arXiv1003.4256B} and semi-analytical/numerical calculations \citep[e.g.][]{Ma2002PhRvL..88u1301M, Zhang2004MNRAS.347.1224Z, HernandezMonteagudo2009MNRAS.398..790H}. Note that previously when $\sigma_8\sim1$ was assumed, the KSZ term was expected to be over an order-of-magnitude smaller than the Compton $y$, but the ratio is closer with the currently favored value of $\sigma_8\sim0.8$. At $l\sim3000$, we find $\Cksz/\Ctsz \approx 0.19$ (adiabatic), 0.23 (standard), 0.29 (low-$\fgas$), and 0.39 (nonthermal20), where the thermal term is evaluated at 146 or 280 GHz (i.e.\ Compton $y$).

Compared to the standard model, the adiabatic model is 20\% larger, the nonthermal20 model is 6\% smaller, and the low-$\fgas$ model is 10\% smaller at $l=3000$. Since the IGM contribution is the same for all of the models by construction, the differences in the power spectra come from the halo contribution alone, and since the models have the same velocity field the differences are entirely due to the halo gas fractions (see Figure \ref{fig:fgasfstar}). Thus, the power spectrum from the adiabatic model has the highest amplitude while the low-$\fgas$ model has the lowest. The nonthermal20 model does not have the lowest amplitude for the KSZ effect, unlike for the TSZ effect.

In our models, the KSZ signal includes scattering with the nonlinear electron distribution up to the epoch of cosmic reionization. In the fiducial case, hydrogen is assumed to be homogeneously reionized at $\zreion=10$. The corresponding Thomson optical depth for electron scattering, $\tauT=0.08$, is consistent with the recent WMAP 7-year results \citep{Komatsu2010arXiv1001.4538K}. For the standard model, we also consider another case where hydrogen is reionized later, at $\zreion=6$, to show the uncertainty due to the poorly constrained reionization history \citep[see][for recent reviews]{Fan2006ARA&A..44..415F, Morales2009arXiv0910.3010M}. Interestingly, we find that the uncertainty due to the halo gas fraction is comparable to the uncertainty due to the reionization history. Thus, in order to constrain the reionization epoch through the KSZ angular power spectrum, it is clearly necessary to understand the nonlinear contribution from collapsed objects. The nonlinear KSZ and TSZ effects can be studied by cross-correlating maps of the CMB with those from galaxy redshift surveys \citep[e.g.][]{Ho2009arXiv0903.2845H, Shao2010arXiv1004.1301S}.

We have neglected to include the contribution from the inhomogeneous reionization epoch. This component is expected to be comparable or even larger than the contribution from the fully reionized universe \citep[e.g.][]{Santos2003ApJ...598..756S, McQuinn2005ApJ...630..643M, Zahn2005ApJ...630..657Z, Iliev2007ApJ...660..933I}. It is especially important to account for this high redshift component when examining the angular power spectrum near the TSZ null frequency. However, at frequencies $\lesssim150$ GHz and $\gtrsim280$ GHz, the KSZ contribution to the SZ angular power spectrum becomes less important as it is expected to be smaller than the current uncertainty in the TSZ term.

\subsection{SZ angular power spectra}
\label{sec:sz}

\begin{figure}[t!]
\center
\includegraphics[width=\hsize]{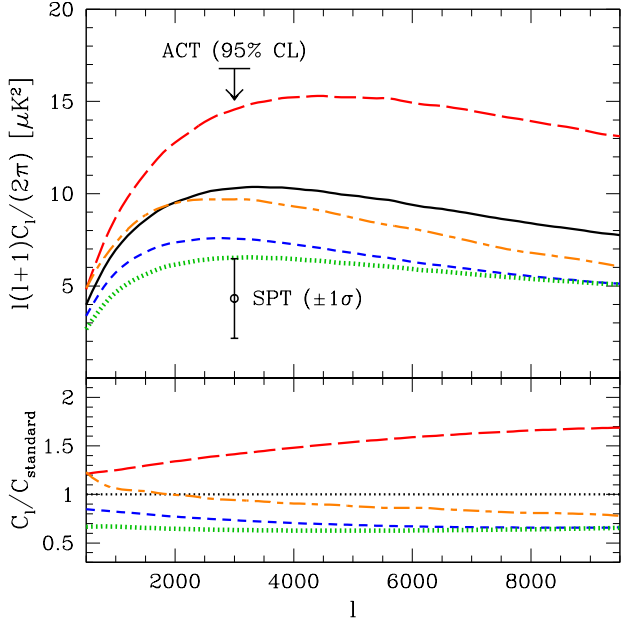}
\caption{{\it Top}: SZ angular power spectra at 148 GHz for the standard (black solid), adiabatic (red long-dash), low-$\fgas$ (blue short-dash), and nonthermal20 (green dotted). The KS02 (orange short-long-dash) template has been scaled to 148 GHz and it does not include any contribution from the KSZ effect. The ACT upper limit is derived using $A_{\rm SZ,STD} < 1.63$ (95\% CL) taken from \citet{Fowler2010arXiv1001.2934T}. The SPT best-fit value is based on $A_{\rm SZ,STD} = 0.42 \pm 0.21$ taken from \citet{Lueker2009arXiv0912.4317L}. This measurement was made at 153 GHz, but the plotted power has been rescaled to 148 GHz. {\it Bottom}: Ratios of power spectra with respect to the standard model.}
\label{fig:cl_sz}
\end{figure}

This section focuses on the frequency-dependent SZ angular power spectrum, mainly near 150 GHz where there are recent measurements by ACT \citep{Fowler2010arXiv1001.2934T} and SPT \citep{Lueker2009arXiv0912.4317L, Hall2009arXiv0912.4315H}. We first compare the models below, then discuss how they scale with $\sigma_8$ in Section \ref{sec:s8}, and finally discuss the constraints from recent observations in Section \ref{sec:discuss}.

Figure \ref{fig:cl_sz} shows our four models and the KS02 model at 148 GHz, which is one of three central observing frequencies of ACT. Relative to the standard model at $l=3000$, the adiabatic model is 41\% larger, the low-$\fgas$ model is 27\% smaller, and the nonthermal20 model is 37\% smaller. The trend in the model amplitudes is the same as found for the TSZ effect, since it contributes several times more power than the KSZ effect at this frequency. In comparison, the KS02 template now has less power than the standard model at $l\gtrsim2000$ because it is based only on the TSZ effect. However, the differences between the two models are $\lesssim10\%$ for the multipole range of current interest ($l\approx3000\pm1000$), where the SZ power begins to exceed that of the CMB but not yet overwhelmed by the rising power from point sources. Thus, the standard and KS02 templates should give similar results when used to fit for the SZ contribution to the CMB angular power spectrum near this frequency.

\subsection{Dependence on $\sigma_8$}
\label{sec:s8}

In order to place constraints on $\sigma_8$, we first need to know how the TSZ, KSZ, and SZ angular power spectra scale with this cosmological parameter. \citet{Komatsu2002MNRAS.336.1256K} suggested that the TSZ power scales approximately as $C_{\rm TSZ} \propto\sigma_8^7$. Using perturbation theory, \citet{Vishniac1987ApJ...322..597V} showed that the large-scale OV angular power spectrum goes as $C_{\rm OV} \propto\sigma_8^4$. With the addition of nonlinear contributions, the KSZ power should have a stronger dependence. To quantify how the SZ templates scale with $\sigma_8$, we define the (q = SZ, TSZ, KSZ) scaling amplitude,
\begin{equation}
A_{\rm q} \equiv \frac{C_{\rm q}(l, \sigma_8)}{C_{\rm q}(l, \sigma_8=0.8)} \equiv \left(\frac{\sigma_8}{0.8}\right)^{\alpha_{\rm q}},
\end{equation}
which is assumed to be a powerlaw relation with $\sigma_8$. In general, the scaling amplitude and the scaling index,
\begin{equation}
\alpha_{\rm q} \equiv \frac{{\rm d}\ln A_{\rm q}}{{\rm d}\ln s},
\end{equation}
where $s\equiv\sigma_8/0.8$, are also dependent on multipole $l$, frequency $\nu$, and astrophysical and cosmological parameters.

We can rescale the four models to arbitrary $\sigma_8$ without having to run additional simulations. Each model is constructed from gas in identifiable halos and in the IGM, and their contributions are rescaled as follows. At $l\sim3000$, gas in identifiable halos (with $\Mfof > 7\times 10^{12}\ \Msunh$ at $z<3$) contribute $\sim95\%$ and $\sim30\%$ to the TSZ and KSZ power spectra, respectively. Since the power from this component is dominated by the one-halo term \citep{Komatsu1999ApJ...526L...1K} and proportional to the halo mass function, we can rescale it with the multiplicative factor,
\begin{equation}
r_{\rm q}(l,\sigma_8) = \dfrac{\displaystyle\sum_{i} C_{{\rm q}, i}(M,z,l,\sigma_8=0.8)w_i(M,z,\sigma_8)u_{{\rm q}, i}^2(\sigma_8)}{\displaystyle\sum_{i} C_{{\rm q}, i}(M,z,l,\sigma_8=0.8)},
\label{eqn:rsz_halo}
\end{equation}
which is calculated by weighting the contribution of each halo $i$ by the ratio of the number density per unit mass,
\begin{equation}
w(M,z,\sigma_8) \equiv \dfrac{\dfrac{{\rm d}n}{{\rm d}M}(M,z,\sigma_8)}{\dfrac{{\rm d}n}{{\rm d}M}(M,z,\sigma_8=0.8)}.
\label{eqn:w_halo}
\end{equation}
We use the \citet{Jenkins2001MNRAS.321..372J} fitting formula for the halo mass function, which agrees well with the simulated halo abundance. The velocity scaling factor $u_{\rm SZ}$ for the SZ power is a linear combination of $u_{\rm TSZ}=1$ and $u_{\rm KSZ}=\sigma_8/0.8$. Gas in the IGM at $z<3$ and all gas at $3 \leq z < 10$ together contribute $\sim5\%$ to the TSZ power, but $\sim70\%$ of the KSZ power at $l\sim3000$. For this low-density component, we assume the linear-regime scaling factors $r_{\rm TSZ} = (\sigma_8/0.8)^2$ and $r_{\rm KSZ} = (\sigma_8/0.8)^4$. Since the power is dominantly from the OV effect, $r_{\rm SZ} \approx r_{\rm KSZ}$ at all frequencies.

\begin{figure}[t!]
\center
\includegraphics[width=\hsize]{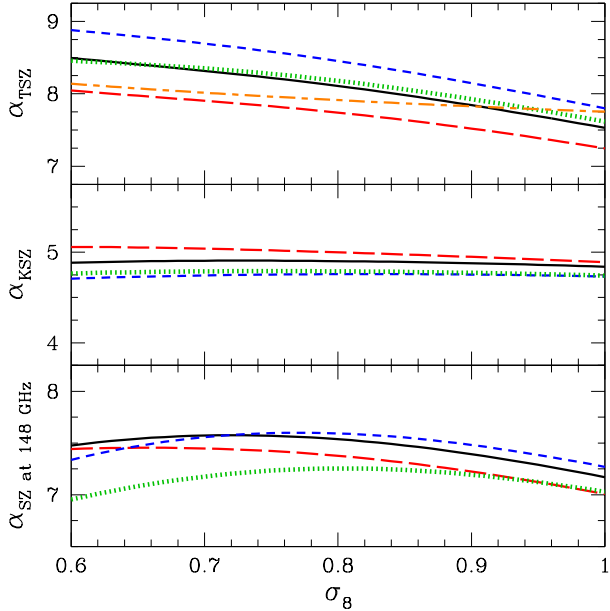}
\caption{{\it Top}:  Scaling index $\alpha_{\rm TSZ}$ for the TSZ angular power spectrum for the standard (black solid), adiabatic (red long-dash), low-$\fgas$ (blue short-dash), nonthermal20 (green dotted), and KS02 (orange short-long-dash) models at $l=3000$. {\it Middle}: Scaling index $\alpha_{\rm KSZ}$ for the KSZ angular power spectrum. {\it Bottom}: Frequency-dependent scaling index $\alpha_{\rm SZ}$ at 148 GHz.}
\label{fig:asz}
\end{figure}

Figure \ref{fig:asz} shows the effective scaling index as a function of $\sigma_8$ at $l=3000$. For the TSZ component, we find $7 \lesssim \atsz \lesssim 9$ for $1 > \sigma_8 > 0.6$; at $\sigma_8\approx0.8$, $\atsz\approx 7.7$ (adiabatic), 8.1 (standard), 8.2 (nonthermal20), and 8.5 (low-$\fgas$). The thermal scaling index increases in models where massive or higher redshift halos make a relatively larger contribution to the power. Considering $\Ctsz$ as the sum of halo terms $C_{{\rm halo}, i}$ and an IGM term $C_{\rm IGM}$, the thermal scaling index can be written as
\begin{equation}
\atsz = \dfrac{{\rm d}\ln\ (x_{\rm IGM}s^2 + \displaystyle\sum_{i}x_{{\rm halo}, i} w_i) }{{\rm d}\ln s}, \end{equation}
where $x_{{\rm halo}, i}\equiv C_{{\rm halo}, i}(s=1)/\Ctsz(s=1)$ and $x_{\rm IGM}\equiv C_{\rm IGM}(s=1)/\Ctsz(s=1)$. The halo abundance scaling $w(M,z,\sigma_8)$, defined in Equation (\ref{eqn:w_halo}), rapidly increases (decreases) with mass and redshift at fixed $\sigma_8$ greater (less) than 0.8. As discussed in Section \ref{sec:tsz}, both massive halos ($M_{500} \gtrsim 3\times10^{14}\ \Msun$) and high redshift halos contribute increasingly larger fractions of the thermal power going from the adiabatic model to the low-$\fgas$ model. This explains why $\atsz$ is smallest for the adiabatic model and largest for the low-$\fgas$ model. At any given mass and redshift, $w$ monotonically increases with $\sigma_8$, but its slope, ${\rm d}w/{\rm d}\sigma_8$, gradually decreases, resulting in the generally observed relation ${\rm d}\atsz/{\rm d}\sigma_8<0$. In all of the models, $\atsz$ at $\sigma_8\approx0.8$ decreases by approximately 0.5 going from $l=3000$ to $l=8000$ since low mass halos become relatively more important at smaller angular scales.

For the KSZ component, we find $\aksz\approx 4.8$ (low-$\fgas$), 4.8 (nonthermal20), 4.9 (standard), and 5.0 (adiabatic) at $\sigma_8\approx0.8$. This trend is different from that for the TSZ effect. Considering $\Cksz$ as the sum of a linear term $C_{\rm L}$ (with $\alpha_{\rm L} = 4$) and a nonlinear term $C_{\rm NL}$ (with $\alpha_{\rm NL} > 4$), the kinetic scaling index can be written as
\begin{equation}
\aksz = \dfrac{{\rm d}\ln\ (x_{\rm L}s^{\alpha_{\rm L}} + x_{\rm NL}s^{\alpha_{\rm NL}})}{{\rm d}\ln s},
\end{equation}
where $x_{\rm L}\equiv C_{\rm L}(s=1)/\Cksz(s=1)$ and $x_{\rm NL}\equiv C_{\rm NL}(s=1)/\Cksz(s=1)$. Since all of the models share the same linear component, $\aksz$ only depends on the nonlinear variables. On one hand, the nonlinear power ratio $x_{\rm NL}$ increases in models with larger gas fractions, being smallest for the low-$\fgas$ model and largest for the adiabatic model, as seen in Figure \ref{fig:cl_ksz}. On the other hand, the nonlinear scaling index $\alpha_{\rm NL}$ has the opposing trend, being lowest for the adiabatic model and highest for the low-$\fgas$ model, as with $\atsz$ in Figure \ref{fig:asz}. In our models, the trend in $\aksz$ is explained by the fact that the differences in $x_{\rm NL}$ slightly outweigh the differences in $\alpha_{\rm NL}$. At $l=3000$, there is weak variation with $\sigma_8$, since the linear component is generally more dominant and $\alpha_{\rm L}$ is independent of $\sigma_8$ (assuming homogeneous reionization). At $l\sim8000$, the larger nonlinear ratio results in $\aksz$ increasing by approximately 0.3 at $\sigma_8\approx0.8$, and there is larger variation with $\sigma_8$.

Figure \ref{fig:asz} also shows $\asz$ at 148 GHz. Since $\atsz$ and $\aksz$ are different functions of $\sigma_8$, $\asz$ will be frequency-dependent. In the nonrelativistic limit, the SZ power is given by $\Csz(\nu) = \Ctsz(\nu) + \Cksz$, and the SZ scaling factor and index can be written as
\begin{align}
\Asz & = x_{\rm TSZ}\Atsz + x_{\rm KSZ}\Aksz,\\
\asz & = \dfrac{{\rm d}\ln\ (x_{\rm TSZ}s^{\atsz} + x_{\rm KSZ}s^{\aksz})}{{\rm d}\ln s},
\end{align}
respectively, where $x_{\rm TSZ}(\nu)\equiv\Ctsz(s=1)/\Csz(s=1)$ and $x_{\rm KSZ}(\nu)\equiv\Cksz(s=1)/\Csz(s=1)$. At $\sigma_8\approx0.8$, we find $\asz\approx7.3$ (nonthermal20), 7.4 (adiabatic), 7.5 (standard), and 7.6 (low-$\fgas$). The trends for $\asz$ are generally similar to those for $\atsz$, because the thermal component contributes several times more power than the kinetic at this frequency. However, the nonthermal20 model is an exception. Even though it has the second largest $\atsz$ amongst the models, it has the lowest $\asz$ because it also has the lowest thermal power ratio $x_{\rm TSZ}$. The SZ scaling index does not vary monotonically with $\sigma_8$, and the relation can be complicated in general. Toward lower $\sigma_8$, $\atsz$ gradually increases, but the lower $\aksz$ increases in weight. Toward higher $\sigma_8$, $\atsz$ gradually decreases, but it also has more weight.

We also calculate $\atsz$ for the KS02 template by repeating the halo model calculation described in \citet{Komatsu2002MNRAS.336.1256K}. At $\sigma_8\approx0.8$ and $l=3000$, $\atsz\approx 7.9$ is higher than the previously suggested value of approximately 7. With the addition of the KSZ component, $\asz$ should be similar to that from the standard and adiabatic models.

\section{Discussion}
\label{sec:discuss}

\begin{figure}[t!]
\center
\includegraphics[width=\hsize]{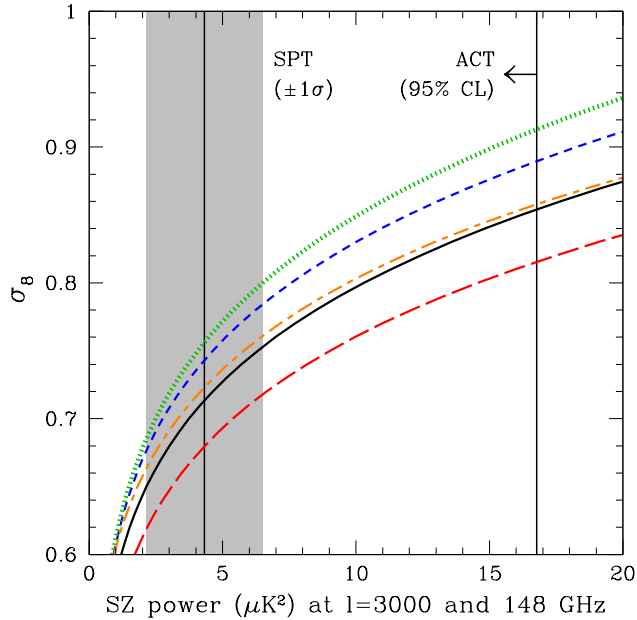}
\caption{Constraints on $\sigma_8$ from the SZ power at $l=3000$ and 148 GHz, assuming the standard (black solid), adiabatic (red long-dash), low-$\fgas$ (blue short-dash), nonthermal20 (green dotted), and KS02 (orange short-long-dash) models. The ACT (95\% CL) upper limit \citep{Fowler2010arXiv1001.2934T} and the SPT best-fit value \citep[][$1\sigma$ uncertainty in gray]{Lueker2009arXiv0912.4317L} for the SZ power are derived using the reported $A_{\rm SZ,STD}$.}
\label{fig:s8}
\end{figure}

ACT and SPT have used our standard model to fit for the SZ contribution to the CMB angular power spectrum on arcminute scales. Figure \ref{fig:cl_sz} shows the constraints on $\sigma_8$ based on the measured $A_{\rm SZ,STD}$, which we convert to an SZ power at $l=3000$ and 148 GHz to compare with the models. The following comparison is based on published SZ data points and not done rigorously by fitting the observed CMB power spectrum.

\citet{Fowler2010arXiv1001.2934T} place an upper limit (95\% confidence level) of $A_{\rm SZ,STD} < 1.63$ based on ACT observations at 148 GHz. This corresponds to  $<16.8\ \muK^2$ at $l=3000$ and 148 GHz. We find $\sigma_8<0.85$ (standard), 0.82 (adiabatic), 0.89 (low-$\fgas$), and 0.91 (nonthermal20). Since the adiabatic model is known to be incorrect, it is safe to omit the constraints from this model. These bounds remain consistent with the most recent value from the WMAP 7-year results \citep{Komatsu2010arXiv1001.4538K}, but they are not highly constraining. With only one of three observing frequencies used in this first analysis, the degeneracy between the SZ and point source contributions to the CMB angular power spectrum can not be broken to obtain best-fit values. Upcoming ACT measurements, which will also include observations at approximately 220 and 280 GHz, will provide much stronger constraints.

\citet{Lueker2009arXiv0912.4317L} report a best-fit value of $A_{\rm SZ,STD}=0.42\pm0.21$ (at 153 GHz) for SPT observations near 150 and 220 GHz. This corresponds to $4.3\pm2.2\ \muK^2$ at $l=3000$ when converted to 148 GHz. We find $\sigma_8 = 0.71^{+0.04}_{-0.06}$ (standard), $0.68^{+0.04}_{-0.06}$ (adiabatic), $0.74^{+0.04}_{-0.06}$ (low-$\fgas$), and $0.76^{+0.04}_{-0.07}$ (nonthermal20). The central values from this first analysis are generally low compared to the WMAP 7-year measurements \citep{Komatsu2010arXiv1001.4538K}, but still within $1\sigma$ for the low-$\fgas$ and nonthermal20 models and within $2\sigma$ for the standard model. \citet{Lueker2009arXiv0912.4317L} find $\sigma_8=0.773\pm0.025$ in a joint analysis with the WMAP 5-year constraints \citep{Dunkley2009ApJS..180..306D}. We expect a higher $\sigma_8$, closer to 0.8, if the same analysis were repeated assuming the nonthermal20 or low-$\fgas$ models instead of the standard model. Upcoming SPT analysis with more survey area and with a third observing frequency of 90 GHz, will provide more conclusive results.

In addition, ACT \citep{Hincks2009arXiv0907.0461H} and SPT \citep{Staniszewski2009ApJ...701...32S, Plagge2010ApJ...716.1118P} have detected the SZ effect directly for massive galaxy clusters. In particular, \citet{Plagge2010ApJ...716.1118P} found that the stacked radial profile from 15 clusters has a very similar shape to that predicted by the best-fit pressure profile from \citet{Arnaud2009arXiv0910.1234A}. Using the best-fit GNFW parameters provided for the individual clusters, we compared their angular profiles against a fair subset ($M_{500}>5\times10^{14}\ \Msun$ and $z<0.4$) of cluster profiles from the standard model, finding good agreement in both shape and amplitude. If we correct the X-ray masses for the assumed nonthermal hydrostatic support, then the SPT cluster profiles are also in good agreement with the nonthermal20 model.

Recently, \citet{Komatsu2010arXiv1001.4538K} analyzed the WMAP 7-year maps and report a significant ($8\sigma$) detection of the SZ effect at the locations of known galaxy clusters from the REFLEX catalog \citep{Bohringer2004A&A...425..367B}. They stacked 742 clusters with $z\leq0.45$ and calculated the average projected SZ angular profile from the CMB maps. The measured signal is found to be approximately 40\% smaller than expectation based on the best-fit pressure profile from \citet{Arnaud2009arXiv0910.1234A}. This discrepancy could be due to the possibility that the GNFW pressure profile derived jointly from X-ray observations and numerical simulations could be overestimating the pressure at $R/R_{500} \gtrsim 1$ or $R/R_{500} \lesssim 0.1$. Unfortunately, we can not fairly compare our models since the stacked, weighted signal depends on detailed properties of the WMAP beam and noise, which are unknown to us. However, the discrepancy may not originate with the adopted pressure profile, but rather with the highly uncertain scaling relations (e.g.\ $R_{500}-T_X$) used to calculate the expected signal.

\section{Conclusion}

We have modeled the SZ effect by postprocessing a simulation of the large-scale-structure of the universe to include gas in dark matter halos and in the filamentary IGM. All free electrons are accounted for up to the epoch of cosmic reionization, which is assumed to occur homogeneously at $z=10$ ($\tauT\approx0.08$). See \citet{Sehgal2010ApJ...709..920S} for additional details and discussion.

We calculate the first-order TSZ (Compton $y$) and KSZ (Doppler $b$) signals, and the frequency-dependent SZ temperature fluctuations. In the latter, the thermal and kinetic components have higher-order relativistic corrections and are nonlinearly coupled. Sky maps with pixel resolution of 0.4 arcminutes are made by tracing through the simulated electron distribution.

Four models for the halo gas distribution are compared to study how gas physics influence the SZ temperature fluctuations and angular power spectra: adiabatic, standard, low-$\fgas$, and nonthermal20. \citet{Bode2009ApJ...700..989B} previously calibrated the standard model to reproduce the stellar and gas fractions and X-ray scaling relations measured from low redshift clusters and groups. Relative to the standard model, the other models differ by approximately $\pm50\%$ (TSZ), $\pm20\%$ (KSZ for $\zreion=10$), and $\pm40\%$ (SZ at 148 GHz) at $l=3000$ and $\sigma_8=0.8$.

We also calculate the dependence of the angular power spectra on $\sigma_8$. The templates for $\sigma_8=0.8$ are rescaled to arbitrary $\sigma_8$ without having to run additional simulations. To quantify the dependence of $C_l$ on $\sigma_8$, a powerlaw scaling relation $C_l\propto(\sigma_8/0.8)^\alpha$ is assumed, where $\alpha=\alpha(l,\sigma_8)$ is the effective powerlaw scaling index. For the collection of considered models, we find $7\lesssim\atsz\lesssim9$, $4.5\lesssim\aksz\lesssim5.5$, and $6.5\lesssim\asz({\rm 148\ GHz})\lesssim8$ for $0.6<\sigma_8<1$ at $l=3000$. Care should be taken when applying the reported scaling indices to other multipoles, frequencies, or to other gas models.

We summarize some interesting features of the SZ angular power spectrum in our models:
\begin{enumerate}

\item The SZ angular power spectrum decreases in amplitude as gas mass and binding energy is removed through star formation, and as gas is pushed out to larger radii by energetic feedback.

\item Nonthermal pressure support can have comparable effects as energetic feedback on lowering the thermal contribution to the SZ power (for reasonable choices of parameters).

\item The broad peak in the power spectrum shifts to larger angular scales as massive or low redshift halos make a relatively larger contribution to the total power.

\item The TSZ scaling index $\atsz(l,\sigma_8)$ increases as massive or high redshift halos make a relatively larger contribution to the power. It also decreases with both $l$ and $\sigma_8$.

\item The KSZ scaling index $\aksz(l, \sigma_8)$ increases with the nonlinear power ratio $C_{\rm NL}/\Cksz$. It increases with $l$, but has only weak variation with $\sigma_8$ for homogeneous reionization.

\item The frequency-dependent SZ scaling index $\asz(l,\sigma_8,\nu)$ increases with the thermal power ratio $\Ctsz/\Csz$. It does not have to vary monotonically with $\sigma_8$ and can be complicated in general.

\end{enumerate}

%\vspace{24pt}
We thank Nick Battaglia, Eiichiro Komatsu, Kavi Moodley, Neelima Sehgal, David Spergel, and Ryan Warne for helpful discussions. HT is supported by an Institute for Theory and Computation Fellowship at the Harvard-Smithsonian Center for Astrophysics. PB is supported by NSF grant 0707731. Computer simulations and analysis were supported by the NASA High-End Computing Program through the Advanced Supercomputing Division at Ames Research Center, the NSF through TeraGrid resources provided by Pittsburgh Supercomputing Center and the National Center for Supercomputing Applications under grant AST070021N, and the TIGRESS high performance computer center at Princeton University, which is jointly supported by the Princeton Institute for Computational Science and Engineering and the Princeton University Office of Information Technology.

\bibliographystyle{apj}
\bibliography{astro}

\end{document}